\documentclass[12pt,preprint]{aastex}
\usepackage{amsmath}
\usepackage{graphics}      
\usepackage{graphicx}

\begin{document}
\newcommand{\kms}{\mbox{km~s$^{-1}$}}
\newcommand{\s}{\mbox{$''$}}
\newcommand{\mloss}{\mbox{$\dot{M}$}}
\newcommand{\my}{\mbox{$M_{\odot}$~yr$^{-1}$}}
\newcommand{\ls}{\mbox{$L_{\odot}$}}
\newcommand{\ms}{\mbox{$M_{\odot}$}}
\newcommand\mdot{$\dot{M}  $}
\newcommand{\Hi}{H{\sc i}}

\title{THE ASTROSPHERE OF THE ASYMPTOTIC GIANT BRANCH STAR CIT\,6}
\author{Raghvendra Sahai}
\affil{Jet Propulsion Laboratory, MS 183-900, California Institute of Technology,
Pasadena, CA 91109}
\email{sahai@jpl.nasa.gov}
\and 
\author{Galen P. Mack-Crane}
\affil{Department of Physics, Occidental College, Los Angeles, CA 90041}

\begin{abstract}
We have discovered two extended half-ring structures in a far-ultraviolet image taken with the GALEX
satellite of the well-known mass-losing carbon star CIT\,6 (RW\,LMi). The northern (southern) ring is 
brighter (fainter) with a diameter of $\sim15'$ ($\sim18'$). These structures most likely represent the astrosphere resulting from the shock 
interaction of CIT\,6's molecular
wind with the Warm Interstellar Medium, as it moves through the latter.
These data provide a direct estimate of the size of CIT\,6's circumstellar envelope that is a factor $\sim$20 larger than previous estimates based on CO millimeter-wave line data. We find that CIT\,6 has been undergoing heavy mass-loss for at least 93,000\,yr and the total envelope 
mass is $0.29\,M_{\odot}$ or larger, assuming a constant mass-loss rate of $3.2\times 10^{-6}$\my. Assuming 
that the shock front has reached a steady-state and CIT\,6's motion relative to the ISM is in the sky-plane, we measure 
the termination-shock standoff distance directly from the image and find that CIT\,6 is moving at a speed of 
about $\gtrsim$39 (0.17 cm$^{-3}/n_{\text{ISM}})^{1/2}\,\kms$ through the interstellar medium around it. However, 
comparisons with published numerical simulations and analytical modelling shows  
that CIT\,6's forward shock (the northern ring) departs from the parabolic shape expected in steady-state. We discuss 
several possible explanations for this departure.

\end{abstract}


\section{Introduction}
The carbon-rich AGB star CIT\,6 (RW\,LMi) is probably the most well-studied carbon star after IRC+10216  that is known to be experiencing 
heavy mass-loss during its Asymptotic Giant Branch (AGB) evolution. 
Such stars eject large quantities of processed material, enriched 
with carbon manufactured in their interiors as a result of 3-$\alpha$ nucleosynthesis, into the interstellar medium 
via extensive dusty molecular winds that operate during the AGB phase. 
At 400 pc, CIT\,6 is somewhat more distant 
than IRC+10216, and has been extensively
observed from radio to optical wavelengths, with a variety of imaging and spectroscopic
techniques. 

The central star is a long-period variable with a period of about 640 days (Alksnis 1995), a bolometric 
luminosity of about $10^4$\,\ls, and an average mass-loss rate of $3.2\times10^{-6}$\,\my~(Zhang et al. 2009) resulting in a 
large circumstellar envelope (CSE) expanding at about 18\,\kms. HST imaging 
at optical and near-infrared wavelengths reveals the presence of a small, roughly bipolar nebula, suggesting that 
the object is transitioning into the pre-planetary nebula phase (Schmidt et al. 2002). These authors also found the 
presence of faint, diffuse arcs $1{''}-4{''}$ from the central star, and suggested that the primary star has 
a main-sequence 
companion of spectral type A-F at a separation greater than 40\,AU. Recently, 
Claussen et al. (2011) discovered the presence of multiple, partial circumstellar ring structures in CIT\,6 at even 
larger distances from the center (up to $\sim8{''}$), from their 
mapping of HC$_3$N J=4-3 emission using the VLA. These arc structures 
have been interpreted and modeled as a 3-dimensional spiral-shock structure 
induced in the CSE due to the presence of a binary companion (Kim et al. 2013). 

The full extent of CIT\,6's CSE has been 
traced most sensitively in CO J=1-0 emission, and has a half-power diameter of $35{''}$ (Neri et al. 1998). 
But since the outer extent of CO emission is limited by the photodissociation of this species by 
the interstellar ultraviolet radiation to $\sim\,2\times10^{17}$\,cm, direct evidence for AGB mass-loss in CIT\,6 is 
limited to a relatively short (expansion) time-scale of about $\sim$3500\,yr. Thus 
the total amount of matter ejected into the ISM by CIT\,6, which depends linearly on the envelope's outer extent 
if the mass-loss has been constant, remains unknown, and likely vastly underestimated. 

In this paper, we report deep GALEX 
images that trace the CIT\,6 CSE to an outer radius that is more than an order of magnitude larger than the above estimate, and likely 
represents the full duration of the current heavy mass-loss. The outer edge of the CSE 
has become visible as a result of its interaction with the ISM as CIT\,6 moves through the latter. We report our
analysis of the shape, size and structure of the CSE-ISM interaction, and compare our results to those expected from 
theoretical models and numerical simulations. We determine CIT\,6's motion relative to the local ISM
and provide new lower limits for the duration of heavy mass-loss and the total mass of ejecta in this object.

\section{Observations \& Results}

\begin{figure}[htbp]
\begin{center}
\includegraphics[width=16cm]{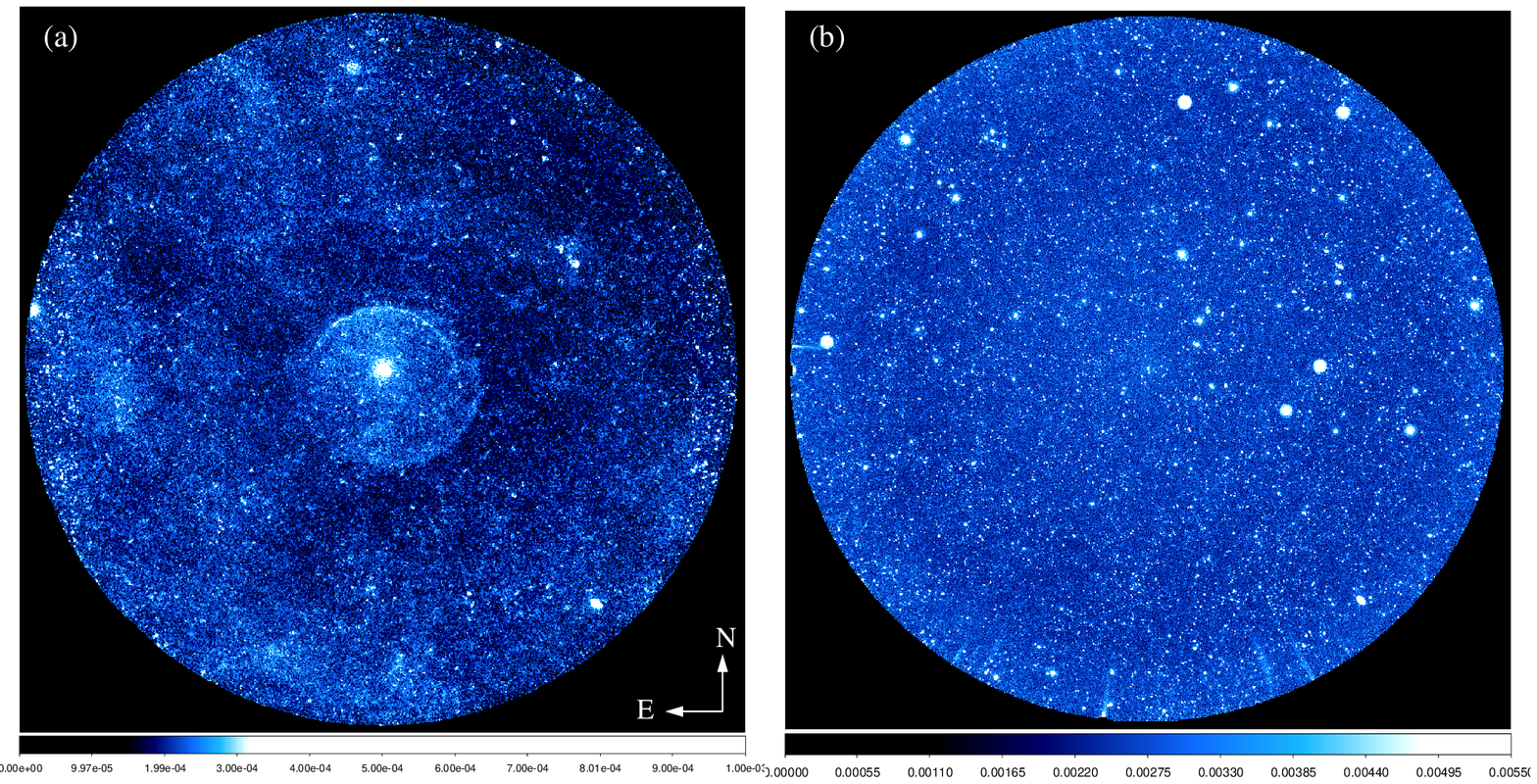}
\end{center}
\caption{(a) FUV GALEX image of CIT\,6 (the circular
field-of-view (FOV) has a diameter of 72.8'$\times$72.8'); the image was boxcar-smoothed
using a $3\times3$ pixel box, and displayed using a linear stretch (in false color). Units in the colorbar are
cps/pixel, implying a flux of $107\,\mu$Jy per $1.5{''}\times1.5{''}$ pixel
(b) NUV GALEX image of CIT\,6 (same FOV as in a), displayed using a linear stretch
(in false color). Units in the colorbar are
cps/pixel, implying a flux of $35.3\,\mu$Jy
per $1.5{''}\times1.5{''}$ pixel
}

\label{fuvnuv}
\end{figure}

We retrieved pipeline-calibrated FUV and NUV images of CIT\,6 from the GALEX archive; the
bandpass  (angular resolution) is 1344-1786\,\AA ($4.5{''}$) and 1771-2831\,\AA ($6.0{''}$), 
respectively, and the pixel size is $1.5{''}\times1.5{''}$ (Morrissey et al. 2005). The data
were taken on
2009 Feb 01, each with an exposure time of 30824 sec. In Fig.\,\ref{fuvnuv}a, we show an
FUV image of the star and its CSE, and in Fig.\,\ref{fuvnuv}b, the corresponding NUV image. Field 
stars in the FUV image have been removed using a customised IDL routine which replaces a small region
covering each star's PSF with a tile of random noise representative of the surrounding sky. The sky
noise was sampled separately at the four corners of each tile and linearly interpolated throughout,
so as to preserve gradients in the local sky background to first order.

\begin{figure}[htbp]
\begin{center}
\includegraphics[width=16cm]{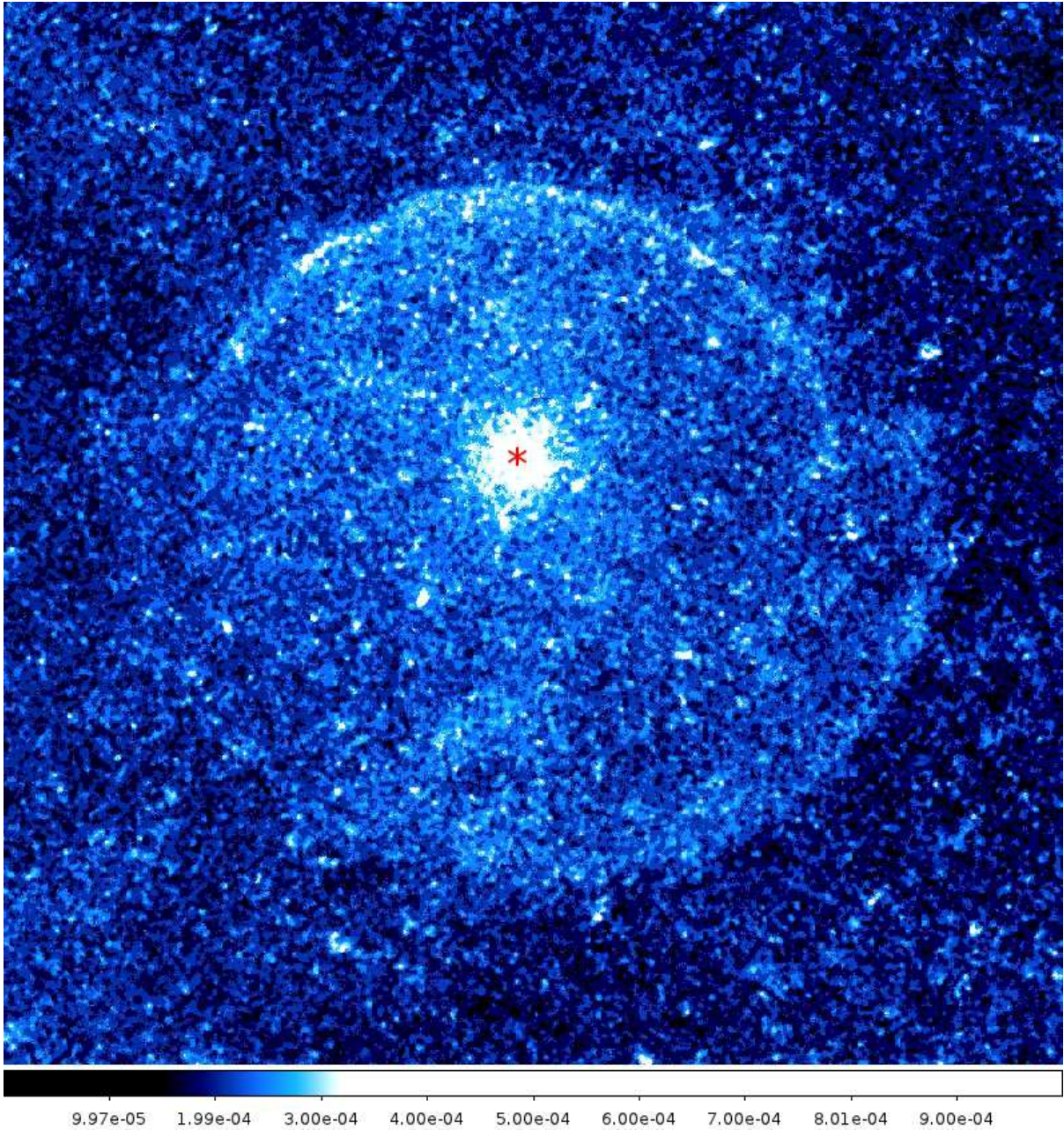}
\end{center}
\caption{As in Fig.\,\ref{fuvnuv}(a), but showing a magnified view of the FUV emission structures around CIT\,6. 
The panel size is 24.75'$\times$24.75'. The location of the central star is marked with a *.
}
\label{fuvbig}
\end{figure}

A magnified view of the relatively bright FUV nebula seen around CIT\,6 is shown in Fig.\,\ref{fuvbig}. 
Bright nebulosity can be seen in the center of the FUV image, around the location of CIT\,6's
central star. In addition, the image shows two bright, extended (size $\sim14'$)
half-ring structures, with diameters of about $15'$ and $18'$. No detectable counterpart to these ring structures 
is found in the NUV image (Fig.\,\ref{fuvnuv}b). 

\begin{figure}[htbp]
\begin{center}
\includegraphics[width=16cm]{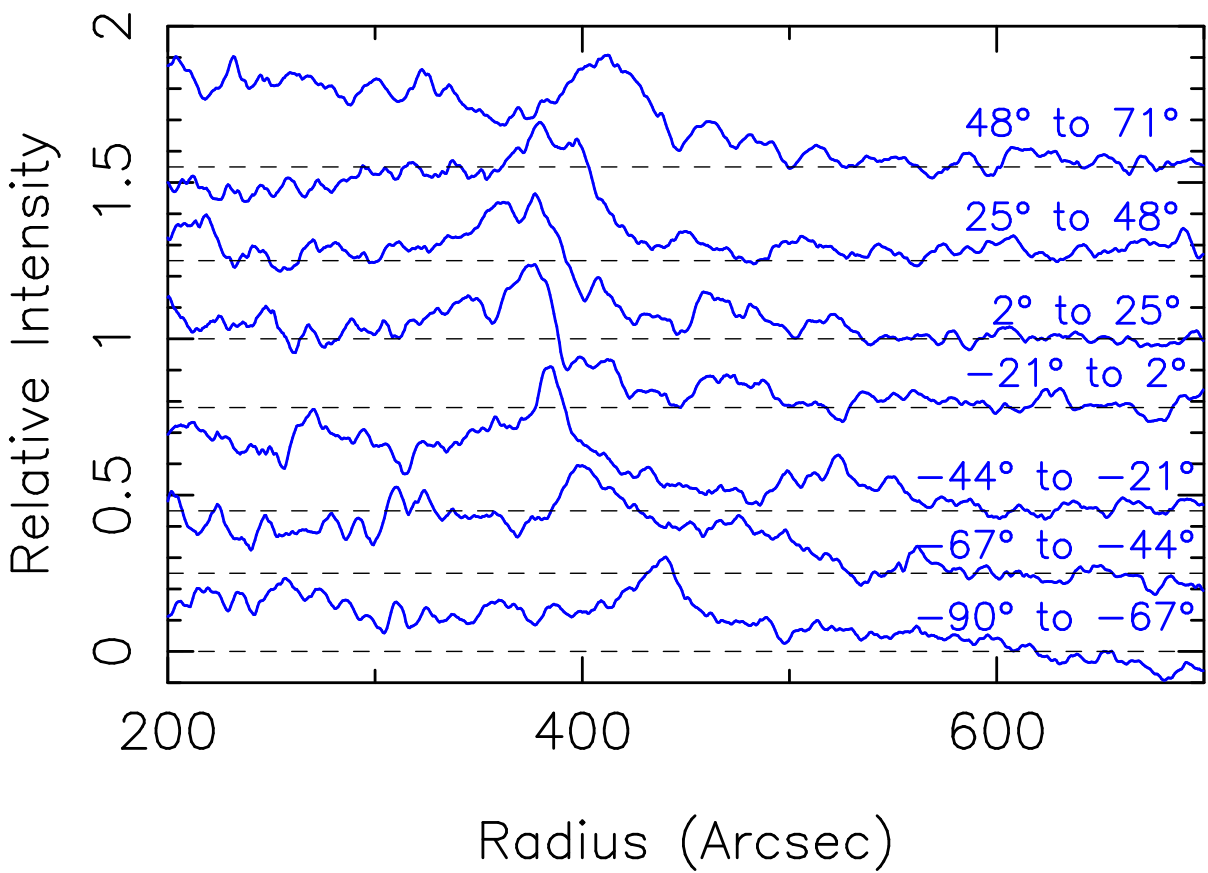}
\end{center}
\caption{Cuts (solid blue curves) of the radial intensity 
averaged over each of seven $23^{\circ}$ angular wedges spanning the northen ring seen in the FUV emission nebulosity around 
CIT\,6. The intensity cuts have been shifted vertically by constant offsets, indicated by
black dashed lines, for clarity.
The cut $PA$s are noted above the cuts on the image.
Intensities are displayed in units of $2.00\times10^{-4}$\,cps/pixel ($0.0214\,\mu$Jy/pixel).
}
\label{wedgeanalysis}
\end{figure}

\begin{figure}[htbp]
\begin{center}
\rotatebox{270}{\includegraphics[width=12cm]{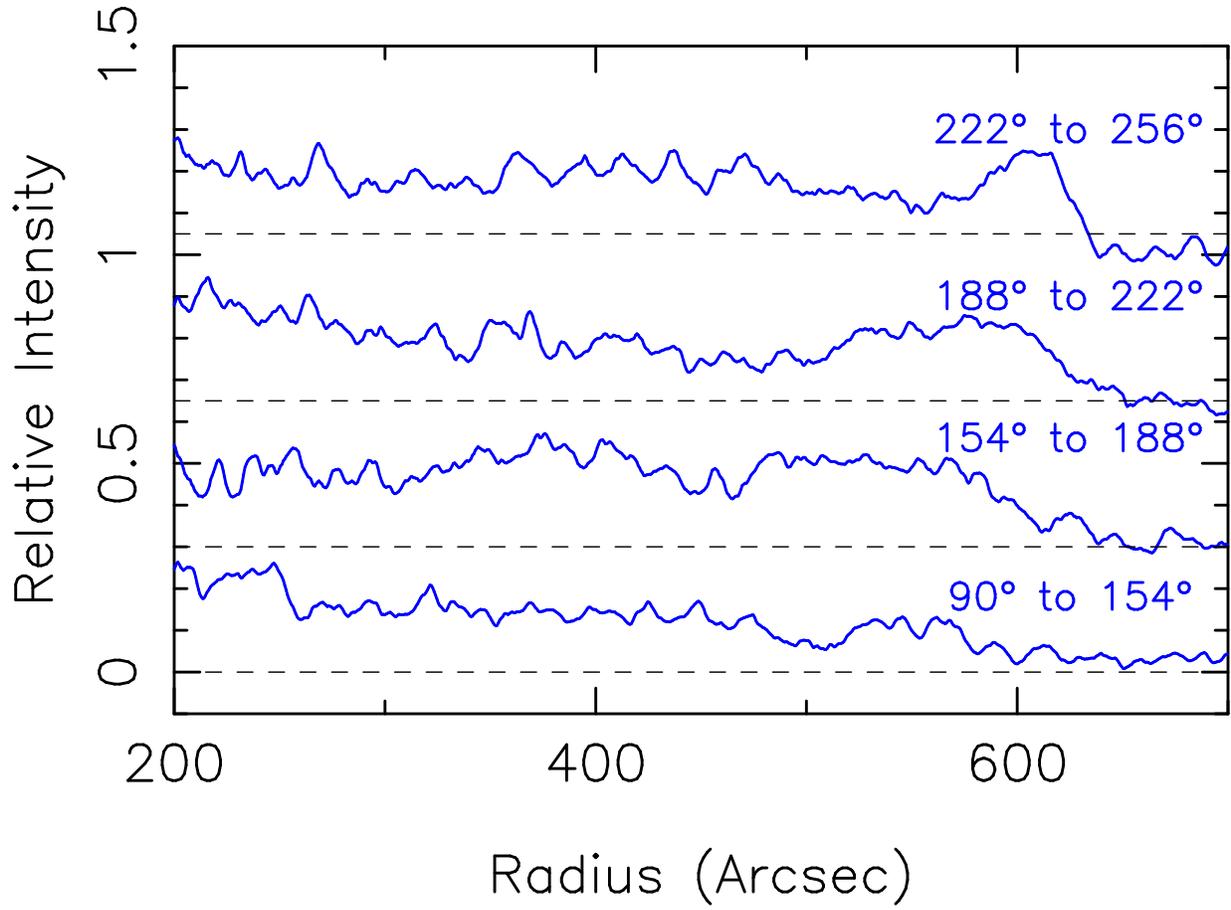}}
\end{center}
\caption{As in Fig.\,\ref{wedgeanalysis}, but for angular wedges spanning the southern ring seen in the FUV emission nebulosity around 
CIT\,6.
}
\label{southring}
\end{figure}

Although the northern ring appears to be roughly circular around the
central star's location, closer inspection supported by examination of radial intensity cuts at 
different position angles (Fig.\,\ref{wedgeanalysis}) shows it be flattened in the northerly
direction. On the southern 
side, the ring is not as prominently limb-brightened as in the north, mostly appearing as a bright 
edge which is roughly circular but at a larger radius from the central star than the northern ring (Fig.\,\ref{southring}). 




\section{An Astrosphere around CIT\,6}\label{structure}
In a large imaging survey at 70 and 160\,\micron~with the Herschel Space Observatory that revealed bow shocks 
for $\sim40$\%, and detached rings for $\sim20$\%, of a sample of 78 evolved stars (AGB stars and red supergiants), Cox et al. (2012: hereafter Cetal12) did not find any wind-ISM interaction structure 
around CIT\,6. However, they did find ``eye" shaped wind-ISM interaction structures (``two elliptical non-concentric arcs at opposing sides of the central source, both have a covering angle of $\lesssim180$\arcdeg") around seven AGB stars that resemble the northern and southern FUV ring structures in CIT\,6. Given the strong indirect evidence that CIT\,6's central star is a 
binary, it is interesting that Cetal12 find that 5 of their 7 ``eye" objects show evidence of binarity\footnote{these 5 include 2 potential binaries and 3 visual binaries: see Cetal12's Table 1 for details and references}. Cetal12 stated that their data were not adequate to either confirm or exclude a connection between
binarity and the ``eyes" morphology. With the inclusion of CIT\,6 to this list, the evidence for such a connection is strengthened.

A plausible interpretation of the FUV emission ring structures is that they  
represent the interaction of the expanding CSE of CIT\,6 with
the local ISM. We consider alternative interpretations later, but find them less likely (\S\,\ref{altmod}). 
The shorter radial distance and greater brightness of the northern ring from the star, 
compared to that of the southern one, implies that the star is moving roughly northwards
through the local ISM, producing a strong shock front at the northern outer edge of
its CSE. 

The northern FUV ring around CIT\,6 thus represents the astrosheath with the outer edge of this ring corresponding 
to the astropause, and the inner
edge to the termination shock (see Fig.\,2d, Ueta 2008). The region interior to the latter
consists of the unshocked, freely-streaming stellar wind; the innermost part of this region is seen
due to the scattering of ambient Galactic starlight from dust in the wind. 

A similar large astrosphere was found 
around IRC+10216 in GALEX images by Sahai \& Chronopoulos (2010: SC10).
Assuming, as for IRC+10216's astrosheath (SC10), that the FUV emission mechanism is due to collisional
excitation of H$_2$ by hot electrons in shocked gas, which 
produces no detectable counterpart in the NUV band (see Martin et al. 2007), the non-detection of the ring 
structure in the NUV is not surprising. We note that the brightest region of the astrosheath 
has an excess FUV intensity (over the background)  
of about $0.013\,\mu$Jy\,pix$^{-1}$, whereas the $3\sigma$ noise in the NUV in this region is $0.036\,\mu$Jy\,pix$^{-1}$.  
Assuming the relative FUV-to-NUV brightness for the astrosheath in 
CIT\,6 is the same as in IRC+10216 (about 6, SC10), the CIT\,6 NUV image, even 
with smoothing to reduce the noise, lacks the sensitivity needed to detect a counterpart to the FUV emission 

We follow a similar procedure as described in SC10 to analyze CIT\,6's astrosphere. 
We measured the northern ring's radial offset in different directions from the central star, using
radial intensity cuts at different position angles. Since the
emission from the ring (the astrosheath) is rather faint, we averaged the intensity over seven
$23^{\circ}$ wedges spanning the northern limb. These cuts (Fig.\,\ref{wedgeanalysis}) show that
the radius of the astrosheath varies systematically, reaching a minimum roughly in the northward
direction, and implying that CIT\,6 motion through the local ISM is indeed northward. 

We have fit a model radial intensity curve derived from a limb-brightened spherical shell to
the FUV radial brightness
profiles, assuming the surface brightness to be proportional to the column density,
and extracted the astrosheath's inner and outer radii ($R_1$ and $R_c$, using the nomenclature in
Fig.\,1 of Weaver et al. 1977). 
We assumed a two-piece inverse-square density profile in our model, 
one for $120{''}\lesssim\,r<R_1$, and the other for $R_c>r>R_1$,
with a jump in density at $r=R_1$. 

\begin{figure}[htbp]
\begin{center}
\includegraphics[width=16cm]{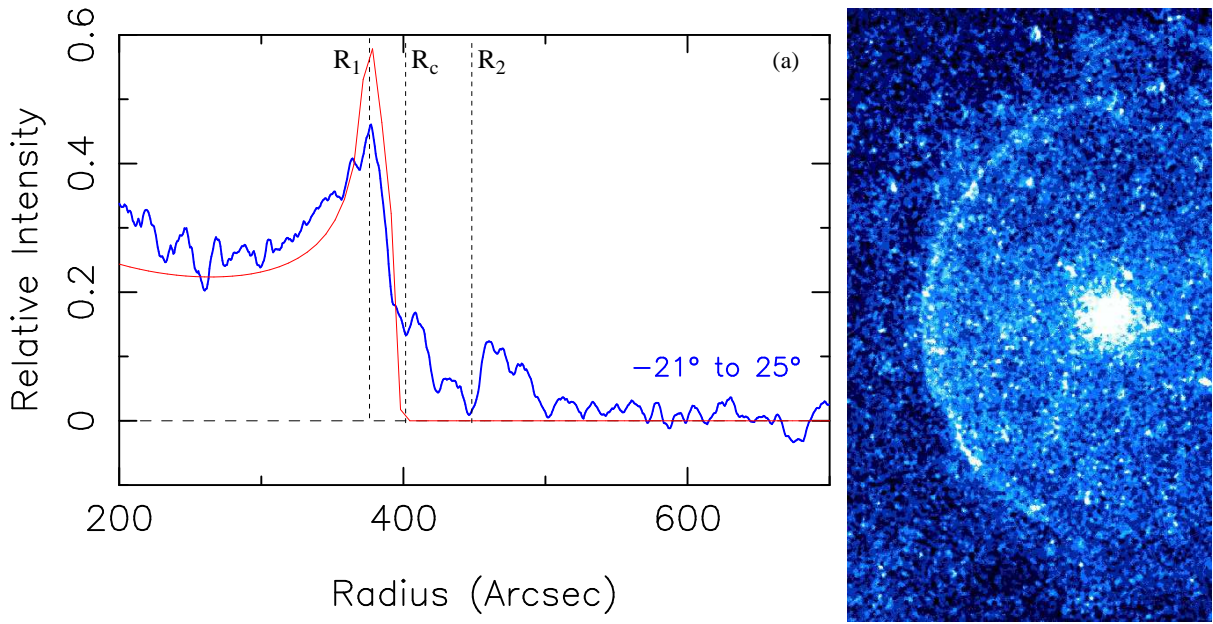}
\end{center}
\caption{(a) Cut of radial intensity (blue) averaged over a 46$^{\circ}$ degree wedge centered at
2$^{\circ}$ in the 
FUV. A model fit to the intensity is shown in red, along with the approximate locations of the
termination shock ($R_1$) and the astropause ($R_c$). FUV intensity units are as in Fig.\,\ref{wedgeanalysis}.
(b) Detail of the FUV emission, rotated from its original orientation (as in Fig\,\ref{fuvnuv}).
Box size is $10.25{'} \times 17.5{'}$; smoothing, stretch, and units as in Fig\,\ref{fuvnuv}. }
\label{sym-axis-fit}
\end{figure}

From a 46$^{\circ}$ wedge centered at $PA=2^{\circ}$ which encloses the symmetry axis 
(Fig.\,\ref{sym-axis-fit}a), we find  $R_1=373{''}$ and $R_c=396{''}$.
These values of $R_1$ and
$R_c$ are not sensitive to the assumed density profiles within these two regions, as they 
are largely determined by the radial location of the intensity peak and the radial width of the 
steeply-falling intensity curve just beyond this peak (e.g., see model fit in Fig.\,\ref{sym-axis-fit}a). 

We cannot derive absolute values of the densities from our
modelling since the proportionality factor between the brightness and the column density is purely
phenomenological; furthermore, since the emission mechanisms in the two regions are different,
the value of the derived density jump is not physical.

We note the systematic presence of a ``shoulder" in the FUV radial intensity cuts, extending to about $50{''}$ beyond
the outer edge of the astropause ($r=R_c$) in the four cuts nearest to the symmetry axis. Although comparably bright features are 
also seen at larger radii beyond this shoulder, their location varies from cut to cut. It is plausible that 
these features have the same origin 
as other patchy emission regions that are present in many other parts of the full field-of-view 
(see Fig.\,\ref{fuvnuv}a), and are likely due to 
scattered light from dust in the ISM, unrelated to CIT\,6. Our tentative conclusion is that the shoulder emission is 
due to a coherent structure that lies just outside the astropause, i.e., $r>R_c$ -- this structure can be seen marginally in  
Fig.\,\ref{sym-axis-fit}b, and probably represents the bow-shock interface separating the shocked and unshocked ISM. A similar 
structure was found by SC10 in IRC+10216. 

The post-shock
temperature in the bow-shock region is expected to be high, about $(3/16\,k$)\
$\bar{\mu}\,V_*^2\sim2\times10^4$K (assuming a strong shock, where the stellar velocity relative to the
ISM, $V_* = 39$\,\kms, \S\,\ref{motion}), where
$\bar{\mu}\sim10^{-24}$\,g is the mean mass per particle for fully ionized gas.
The emission in this region is thus most likely dominated by
the two-photon continuous emission of H (Spitzer \& Greenstein 1951).


\section{CIT\,6's Motion through the ISM, Mass-Loss Duration and Circumstellar
Mass}\label{motion}
We estimate the star's velocity $V_*$ through the surrounding ISM using the relationship between
$l_1$, the distance of the termination shock from the star along the astropause's symmetry axis
(i.e., the termination-shock standoff distance), and $V_{*} (\kms)=10\,V_{*,6}$
(Eqn. 1 of van Buren \& McCray 1988\footnote{there is a missing minus sign in the exponent of $\bar{\mu}_H$ in their equation, which we have corrected below}):
\begin{equation} l_1 (\text{cm}) =1.74\times10^{19}\,(\dot{M}_{*,-6} V_{w,8})^{1/2}\
(\bar{\mu}_H\,n_{\text{ISM}})^{-1/2}\,V_{*,6}^{-1}
\label{vanburen} 
\end{equation}

\noindent
where $\dot{M}_{*,-6}$ is the stellar mass-loss rate in units of $10^{-6}$\,\my, $V_{w,8}$ is
the wind velocity in units of $10^3$\,\kms, $\bar{\mu}_H$ is the dimensionless mean molecular mass
per H atom, and $n_{\text{ISM}}$ is the ISM number density in cm$^{-3}$.

Given the strong asymmetry between the northern and southern hemispheres, we first make the
simplifying  
assumption that the astropause's symmetry axis lies in the sky-plane, i.e., the inclination
angle, $\phi=90^{\circ}$. 
We find $l_1=R_1\,D=2.2\times10^{18}$\,cm, using the value of
$R_1=373''$ derived earlier, and the distance D=400\,pc. Substituting this
value of $l_1$ in Eqn.~\ref{vanburen}, with $\dot{M}_{*,-6}=3.2$, $V_{w,8} =
0.018$, and $\bar{\mu}_H = 1.33$ (for an 89/11 mixture of H/He), we find 
$V_* = 39\;(n_{\rm ISM}/0.17)^{-1/2}$\,\kms. Our choice of $n_{\text{ISM}}$
is discussed in \S\,\ref{ism}. The value of $V_*$ (i) does not
depend on the (uncertain) distance, $D$, to CIT\,6, since both $l_1$ and $\dot{M}_{*,-6}^{1/2}$ scale
linearly with $D$, and (ii) depends only weakly on the uncertain value of the ISM density at
CIT\,6's location. The inclination angle $\phi$ may be significantly smaller than $90^{\circ}$ -- we 
discuss this in \S\,\ref{shap}.

\subsection{Density of the ISM around CIT\,6}\label{ism}
We estimate the ISM number density near CIT\,6 based on the star's location in the Galaxy as follows. 
First we determine the the density of neutral hydrogen, H{\sc i} around CIT\,6. 
\noindent We approximate the H{\sc i} disk scale height $h_z (R)$ using the relation (Kalberla \& Kerp 2009)
\begin{equation} h_{z}(R)=h_0 e^{(R-R_{\odot})/R_0},\; \text{for}\; 5<R<35 \,\text{kpc,}
\end{equation}

\noindent where $R$ is the Galactocentric radius, and $R_{\odot}$ is the distance of the Sun from the Galactic center, 
and $h_0=0.15 \,\text{kpc}$, $R_0=9.8 \,\text{kpc}$. 
Taking $R_{\odot}=8.33 \,\text{kpc}$ (Gillessen et al 2009), and using CIT\,6's galactic coordinates 
($l=197\arcdeg.7147$, $b=+55\arcdeg.9642$, its distance from the 
Sun of $0.4\,\text{kpc}$, we find $R=8.44\,\text{kpc}$, which implies $h_{z}(8.44)=0.152\,\text{kpc}$.
The midplane density of H{\sc i} (i.e., at $z=0$) is found from the relation (Kalberla \& Kerp 2009)
\begin{equation} n_{\text{H\sc i}} \sim n_0 e^{-(R-R_{\odot})/R_n}\; \text{for}\; 7<R<35 \,\text{kpc,}
\end{equation}

\noindent where 
$n_0=0.9$ cm$^{-3}$ is the density at $R=0$ and $R_n=3.15 \,\text{kpc}$. Combining the midplane H{\sc i} density at CIT\,6's Galactocentric radius, $n_{\text{H\sc i}} \sim 0.87$, with 
the fractional H{\sc i} density of 0.038 (given by $e^{-ln(2)(z/h_{z})^2}$) at CIT\,6's 
height above the galactic plane ($z=0.33\,\text{kpc}$), we find a relatively low value for the H{\sc i} density, 
$n_{\text{H\sc i}} \sim 0.033$. Thus it is likely that the ISM
surrounding CIT\,6 is ionized and that CIT\,6 is embedded in the Warm Ionized Medium (WIM), specifically its 
thick-disk that extends more than a kpc above and below the Galactic plane.
The volume-filling factor of the WIM in the midplane in this region is $f=0.04 \pm 0.01$, 
which grows exponentially with $z$ over the $0-1.4 \,\text{kpc}$
range as $e^{\left | z \right |/H_f}$, where the scale height $H_f \sim 0.7 \,\text{kpc}$.  (Gaensler et al. 2008).  
In the neighborhood of CIT\,6, $f(z=0.33\,\text{kpc})=0.065$.
Because the volume-filling factor is small, $n_{\text{ISM}}$ is likely less than 
$n_{\text{typ}}$, where $n_{\text{typ}}$ is the typical internal electron density for clouds in the thick-disk component of the WIM.
In the midplane, $n_{\text{typ}}=0.34 \pm 0.06 \,\text{cm}^{-3}$ and 
decays with $z$ as $e^{-\left | z \right |/H_N}$, where the scale height $H_N=0.5 \,\text{kpc}$ (Gaensler et al. 2008). 

In the neighborhood of CIT\,6, $n_{\text{typ}}(z=0.33\,\text{kpc})=0.17 \,\text{cm}^{-3}$.
Assuming that CIT\,6 is embedded in a WIM cloud, 
$n_{\text{ISM}}=0.17 \,\text{cm}^{-3}$, which is the value that we use for our estimate of $V_*$.
Our value of $n_{\text{ISM}}$ is larger than an estimate by Cetal12, who find
$n_{\text{ISM}}\approx 0.04\,$cm$^{-3}$, assuming a representative global average of the 
density given by $n_H(z)=2 e^{(-\left | z \right |/100pc)}$, although acknowledging that the structure of the ISM 
entails fluctuations in the density (and filling factor) on all spatial scales. 

\subsection{Duration of Mass Loss and Circumstellar Mass}\label{massloss}
The FUV emission traces the AGB stellar wind in CIT\,6 to a much larger distance from the star than
that derived from previous measurements of the CO J=1-0 and 2-1 emission (Neri et al. 1998). These authors 
take the CSE outer radius to be that at which CO is photodissociated by the interstellar radiation field. Since the photodissociation radius depends on 
${\dot{M}}^{0.58}$ and Neri et al. (1998) use \mdot=$7.2\times 10^{-6}\,\my$, we scale their 
value of this radius for our mass-loss rate \mdot=$3.2\times 10^{-6}\,\my$ to derive a photodissociation radius of $1.7\times 10^{17}$\,cm.
This estimate of the CSE outer radius implies a mass-loss duration of 3500\,yr.
  
We use the astropause size to substantially revise (upwards) the above estimate of the
duration, $P$, of heavy mass-loss in CIT\,6. We take the radius  
of the termination shock in the direction orthogonal to the symmetry axis ($432''$ or $2.6\times 10^{18}$\,cm at $D=400$\,pc) 
as a measure of the outer radius to which the unshocked wind has expanded, since  
the radial extent in that direction is independent of the inclination angle. 

We estimate $P$ by deriving expansion time-scales ($P_u, P_s$) for the unshocked and shocked wind regions  
separately; $P_u=45,240$\,yr from the ratio of the termination shock radius to
$V_w$, and $P_s=47,950$\,yr from the ratio of the astrosheath width ($38{''}$)
to an average velocity for this
region, $\bar{V_s}=1.50\,\kms$. We take $\bar{V_s}=V_s/2$, where $V_s=V_w\
(\gamma-1)/(\gamma+1)=V_w/6=3.00\,\kms$, is the velocity in the astrosheath just beyond the
termination shock, with $\gamma=7/5$ for diatomic gas, and assuming the latter to be adiabatic. The
actual value of $V_s$ should be less 
than the adiabatic value, since the astrosheath appears to have cooled to some degree -- the
astrosheath's width of $38{''}$, 
or $2.3\times 10^{17}$ cm, derived from fitting the radial intensity cut in a direction orthogonal to the symmetry axis,
is smaller than the adiabatic
value,  $\approx0.47\,l_1=1.0 \times 10^{18}$ cm (Eqn. 2, Van Buren \& McCray 1988) by factor of about 4. 
Furthermore, once a complete balance has been established
between the ram pressure of the stellar wind and that of the ISM, the leading edge of the astropause (i.e., the forward shock structure)
remains a fixed distance ahead of the moving star (Weaver et al. 1977).
Hence, if the shock interaction of CIT\,6's wind with the ISM has reached equilibrium,   
$P=P_u+P_s=93,190$\,yr is a lower limit, in which case CIT\,6
has been undergoing mass-loss for at least 93,000\,years, and the total CSE mass is
$>0.29\,M_{\odot}$. 

Model fits to IRAS far-infrared scan data have been used to derive dust shell sizes for a 
large sample of AGB stars, including CIT\,6 by Young et al. (1995: Yetal95) -- for the latter, 
they estimated an outer radius of $7\arcmin$. However, this 
result must be regarded with caution as (a) inspection of the data and model fits by Yetal95 for CIT\,6 (see their Fig 7, panels D(i) and D(ii)), as well as many other stars in their sample, shows the presence of large negative and positive fluctuations in the derived intensity profile of the extended dust-shell emission, (b) the Herschel PACS imaging data in the Cetal12 study did not reveal an extended dust shell around CIT\,6, and (c) a comparison of the outer radii of the dust shells for a sample of AGB stars in Yetal95 with those reported to have bow-shock (Class I), 
eye (Class II), or ring (Class III) structures in Cetal12, shows that the IRAS-based value is significantly larger (i.e., by factors as large as $\sim2-5$). 

It is possible that the shells inferred from the IRAS data represent circumstellar structures that lie beyond the dust structures found with Herschel, but were not detected by the latter because of inadequate sensitivity. For example, in the case of Y CVn, 
an extended dust shell is clearly seen in ISO maps (Izumiura et al. 1996) with about the same size as inferred by Yetal95, but only very faintly in the Cetal12 study. However, for CRL2688, where Speck et al. (2000) reported the presence of two dust shells from ISO scan data, Spitzer imaging by Do et al. (2007) showed no such shells at a level well below the intensities expected from the ISO results. We believe that a detailed re-investigation is needed to resolve such discrepancies in results related to the presence of extended dust shells as derived from IRAS and ISO, and from the more recent Spitzer and Herschel missions (utilizing modern large-format detector arrays).

If, however, the much larger circumstellar shells inferred by Yetal95 are real, then their presence 
poses a problem for each such object common to the Cetal12 study in which the (smaller) dust shell represents a bow-shock structure due to the wind-ISM interaction resulting from the object's motion through the ISM (based on its morphology, i.e., Class I and possibly Class II). In order to resolve this problem, one would require the ISM to be streaming  
through ``holes" in the outer IRAS-detected shell in order to interact with the inner circumstellar shell in these 
objects -- numerical simulations would be needed to determine if this is a plausible scenario. 

\subsection{Forward Shock Shape and Inclination}\label{shap}
Has CIT\,6's forward shock structure reached a steady state? The shape of this structure is expected to change with time, being initially circular 
and becoming increasingly parabolic until it reaches its steady state morphology (Weaver et al. 1977), given by the analytic solution of Wilkin (1996). 
We define $\eta(0/90)$ as the ratio between the radius of the termination shock along the shock's symmetry axis to that 
along an orthogonal direction. Employing Eqn. 9 in Wilkin (1996), 
the analytic value is $\eta(0/90)=1/3^{1/2}=0.577$, and is reached after $\sim40,000$ yr 
in simulations of astrospheres around AGB stars that are applicable to CIT\,6 (see Fig.\,11 of Mohamed et al. 2012 [MML12]). 
However, from CIT\,6's observed shock structure, we find 
$\eta(0/90)=373/432=0.86$, assuming that the inclination angle is, $\phi=90^{\circ}$. 

The discrepancy between the observed and expected steady-state 
value of $\eta(0/90)$ in CIT\,6 is not unprecedented -- a similar discrepancy exists 
for the forward shock structures seen in IRC+10216 and the 
red supergiant Betelgeuse, where this parameter is  $\sim0.8$ (SC10, Decin et al. 2012).

In the MML12 simulations, $\eta=0.86$ corresponds to an age of only $\sim5,000$ years. 
We found earlier that CIT\,6 has been undergoing mass loss for a period at least as long as $P_u=45,240$\,yr, and possibly much longer. If it has been 
interacting with the ISM during most of this period it is likely to  
have reached a steady state, in which case the discrepancy between the observed and expected values of $\eta(0/90)$ would 
imply that the star's direction of motion of the star is not in the
plane of the sky.

Mac Low et al.'s (1991: MLetal91) 
paraboloidal shock models for various inclination angles (their Fig.\,5)
show that as $\phi$ becomes {\it smaller}, the ratio of the radial distance between the
star and the
apex of the projected emission paraboloid, to the (unprojected) standoff distance, becomes
{\it larger}. 
A visual comparison of CIT\,6's FUV emission morphology (Fig.\,\ref{fuvnuv})
with the surface brightness contours in Fig.\,5 of MLetal91, indicate that $\phi$ is small,
$\sim30^{\circ}$, and our measured value of $R_1$ is larger than $l_1$ by a significant
factor. We estimate that this factor may be as large as $\sim1.5-2$, by comparing the length of
the unprojected standoff distance vector, to its apparent value as estimated from the distance between the star and 
the apex defined by the emission contours for $\phi=30^{\circ}$ and $90^{\circ}$ in Fig.\,5 of MLetal91.
Thus the value of $V_*$ may be as high as $\sim70$\,\kms.

The value of $\phi$ depends on the orientation of the relative motion vector between CIT\,6 and the WIM cloud with which it is interacting. Thus, if both CIT\,6 and the WIM cloud have similar radial velocities, then the relative motion vector would lie in the sky plane, i.e., $\phi \sim 90^{\circ}$. CIT\,6's radial velocity is 
$V_{lsr}=-1.5$\,\kms~(Olofsson et al. 1993). For the WIM, Haffner et al. (2003) conclude, using data from the Wisconsin H$\alpha$ (WHAM) survey that the H$\alpha$ emitting gas at high galactic latitudes is clearly biased towards negative $V_{lsr}$ values; inspection of the longitude-velocity plots 
for b=50$^{\circ}$ and $60^{\circ}$ in Fig. 9 of their paper shows relatively strong emission extending to 
$V_{lsr}$ values of $-50$ to $-75$\,\kms~at CIT'6's longitude. Thus, it is quite possible that the relative motion vector between CIT\,6 and the local WIM has a significant radial component, implying that $\phi$ is significantly less than $90^{\circ}$.


Alternatively, the density of the ISM around CIT\,6 is much lower than $few\times0.1$\,cm$^{-3}$, or its mass-loss rate was significantly higher in the distant past (since the time required to reach a steady state varies as ($n_{ISM}$/\mdot)$^{1/2}$, Weaver et al. 1977). 
It is also possible that 
CIT\,6 has only recently entered the higher-density region with which it is 
interacting -- this alternative finds some support 
in the relatively smooth shock structure observed in CIT\,6, as hydrodynamical instabilities take time to develop.
For example, MML12 attribute the smoothness of the shock structure of Betelgeuse to the bow-shock interaction being too young for
instabilities to have developed (they constrain the age to $<30,000$\,yr). A note of caution 
here is in order -- the production of instabilities in 
different simulation studies varies significantly, both in terms of the 
time-scales over which they are produced and in their size and structure, 
as noted by Decin et al (2012) while comparing their simulations of 
the bow-shock around the red supergiant Betelguese with those of MML12 and Cetal12. 

Other explanations for a smooth shock structure are also possible, including the suppression 
of large scale instabilities due to a warm ISM (Decin et al 2012) (which may be appropriate for CIT\,6), and the presence of a magnetic field in the 
direction of motion (van Marle et al. 2014).

\subsection{Comparison with IRC+10216}\label{comparison}
It is instructive to compare the astrosphere of CIT\,6 with that of IRC+10216, the only other carbon-rich star that has been found to 
show such a structure in GALEX images (SC10). SC10 noted that IRC+10216's forward shock structure is relatively smooth and does not show the large-scale instabilities expected in such interactions (Blondin \& Koerwer 1998), and CIT\,6 is similar in this respect. 
While the absence of large-scale instabilities for CIT\,6's shock may be explained as a result of it being surrounded by a warm ISM, this 
explanation is not applicable for IRC+10216, which (as we show below) is most 
likely immersed in neutral (and cooler) gas. 

Given IRC+10216's Galactocentric radius $R=8.36$\,kpc (derived from its galactic coordinates $l=221\arcdeg.4466$, $b=+45\arcdeg.0604$, 
and D=120\,pc), we find the H{\sc i} scale height 
to be $h_z = 0.15$\,kpc, giving a fractional density of $0.80$ at its height above 
the Galactic plane of $z=85$\,pc. With a mid-plane density of 0.89\,cm$^{-3}$ at $R=8.36$\,kpc,  
the density of H{\sc i} near IRC+10216 is then
$\sim 0.72$\,cm$^{-3}$, implying that the ISM near IRC+10216 is probably largely neutral (supporting SC10's 
choice of $n_{\text{ISM}} = 1$\,cm$^{-3}$), thus significantly cooler than the WIM. 

The morphologies of the astrotail regions of CIT\,6 and IRC+10216 differ markedly, with CIT\,6 showing a relatively smooth arc structure, whereas IRC+10216 shows prominent vortices. SC10 attribute these as resulting from the vortex shedding seen in numerical simulations of AGB wind-ISM interactions that occurs at very long interaction time-scales ($\sim$500,000\,yr) (Wareing et al. 2007). Thus it appears that the period of wind-ISM interaction in IRC+10216 is 
much longer than that of CIT\,6.


\section{Comparison with 3D Numerical Simulations}\label{model}
We compare our observations of CIT\,6 with the results of 3D simulations 
of the bow-shock around Betelgeuse ($\alpha$\,Ori) by MML12, since there is reasonable similarity 
between the physical parameters relevant to the shock structure employed in these models and those applicable to CIT\,6. 
MML12's model D has $\dot{M}_{*}=3.1\times10^{-6}$\,\my, $V_w=17$\,\kms, 
$V_*=72.5\,\kms$ and $n_{\text{ISM}}=0.3$. The density structure for model D (rightmost panel of Fig.\,7 in MML12)
appears quite similar to our image of CIT\,6 -- specifically, the density plot shows the presence of 
a high-density semi-circular structure in the direction of motion (leftwards in this figure), and a  
roughly semi-circular ``edge" structure in the opposite 
direction. The latter is at a larger radial distance from the central star than the former, similar to what 
we observe in CIT\,6, and shows a density contrast of a factor 10 or more relative to the immediate environment 
beyond it (seen in red/brown hues). We note that the parabolic-shaped 
structure seen in yellow in MML12's Figure 7 represents high-temperature ($>10^4$\,K) ionized gas in the bow-shock, which at best is only seen very faintly, and only around the forward shock in our observations -- hence its absence in the tail region of the 
observed astrosphere of CIT\,6 
is most likely due to inadequate sensitivity.
The southern ring, like the northern one, is only seen in FUV emission and not in the NUV, suggesting that the emission 
mechanism for the former is the same as the latter, namely, collisional excitation of H$_2$ by hot electrons.

MML12 find that the gas in the ``bow shock arc (where the gas is strongly decelerated)", i.e., the astrosheath in our description, 
behaves nearly isothermally, consistent with our finding for CIT\,6 that the average
astrosheath width is significantly smaller than the expected adiabatic value (\S\ref{massloss}).
But as noted earlier, in the simulation  
the stellar wind-ISM interaction has been ongoing 
for 32,000\,yr and is close to steady state with $\eta(0/90)\sim0.6$, whereas the observed value of $\eta(0/90)$ is significantly higher.

\section{Alternative Models for the FUV Rings around CIT\,6}\label{altmod}
We now consider alternatives to the astrosphere model for the origin of the FUV emission rings around CIT\,6. Given the presence of the partial arcs seen by Claussen et al. (2011) and Schmidt et al. (2002) within 
$\sim$8\arcsec~of the central star, believed to be created by a spiral shock due to the presence of a central binary, one possibility is that the much more distant FUV emission rings also have the same origin. In this model, the radial pitch of the spiral, i.e., the separation between successive windings is equal to $P\,V_{exp}$, where $P$ is the binary period 
and $V_{exp}$ is the expansion velocity of the CSE (e.g., Mauron \& Huggins 2006). Hence, assuming $V_{exp}$ does not change substantially over the timescales corresponding to the rings, the current separation of the windings ($\sim2\farcs6$), provides an upper limit to the separation in the past (because the period P of the binary increases as the central star loses mass via its wind). Clearly, given 
the much larger observed difference in the average radius of the northern and southern rings ($\sim200\arcsec$), these cannot represent successive windings, even if we allowed for a $V_{exp}$ that was larger in the past by a factor 10 over its current value.

Another possible model is one in which the northern and southern rings represent two discrete mass-loss episodes in CIT\,6's distant past. However, such a model would then require that each of these episodic ejections was confined to only a hemispherical region -- with either both ejections occcuring at different times with roughly the same velocity, or at the same time with different velocities (or some combination of these). Such a mass-ejection history would be unprecedented for an AGB star, and would not fit into the standard model of roughly spherical AGB mass-loss (i.e., driven by radiation pressure on dust grains). It also does not provide an explanation for the systematic increase in the radius of the northern ring around its symmetry axis. We note that although the so-called detached shell phenomenon observed in a few carbon-rich stars does produce large single shell structures (e.g., the Class III objects in the Cetal12 survey), these shells, believed to result from a relatively short episode of enhanced mass-loss rate (and possibly wind velocity as well) initiated by a He-shell flash (e.g., Olofsson et al. 1990, Steffen \& Sch{\"o}nberner 2000), are circular and complete and thus morphologically different from CIT\,6's FUV emission rings.

\section{Concluding Remarks}
In summary, the GALEX images of CIT\,6 show an unusually detailed picture of the interaction of the wind from a carbon-rich 
AGB star with the ISM due to its motion in the latter. This interaction process has been studied using UV and far-infrared observations for other AGB stars  (e.g., SC10, Cetal12 and references therein)  
and a red supergiant ($\alpha$\,Ori: Ueta et al. 2008, Decin et al. 2012, le Bertre et al. 2012), as well as HI 21\,cm line observations (Matthews et al. 2013, and references therein). The far-infrared imaging observations with Herschel by Cetal12 have discovered the largest number of astrospheres by far, clearly tracing the detailed geometry of the shock front in the wind-ISM interaction for a large sample of AGB stars. The HI observations reveal, for a smaller sample, that the CSEs have been significantly influenced by the wind-ISM interaction, but since the angular resolution of these observations is relatively low ($\sim1\arcmin$), it is difficult to spatially separate the detailed shape and structure of the astropause from the HI emission of the CSE as a whole. However, kinematical information extracted from the observed HI line profiles provides support for deceleration of the freely expanding CSE by the ISM.

Including CIT\,6, there are now three examples of AGB stars (IRC+10216, Mira, and CIT\,6) and one RSG ($\alpha$\,Ori) that show astrospheres in FUV emission. An exhaustive search of the GALEX image archive in order to determine if there are additional such examples would be useful, but is outside the scope of this paper. However, we speculate that such a search would not be very succesful given the very long exposure times that were needed for the GALEX imaging to reveal the relatively faint FUV emission from the wind-ISM interaction in these objects 
(e.g., 30,824\,s for CIT\,6; 8783\,s for IRC+10216; 48,915\,s for $\alpha$\,Ori.)

Our study provides valuable new data for understanding the
CSE-ISM interaction, and shows that there are specific features in the resulting shock structures that are not 
adequately reproduced by current numerical simulations. These features include the relatively high value of 
the parameter $\eta(0/90)$ compared to the steady-state value, and the relative smoothness of the shock structure implying the general lack of large-scale instabilities. 
Given that such discrepancies are seen in other objects as well, new modelling 
that is focussed on addressing these will help improve our understanding of this interaction, and our ability to make inferences about the unprecedented long history of AGB mass-loss revealed by these interactions. For example, one avenue that merits detailed investigation is when the wind-ISM interaction involves an encounter of the star's CSE with a relatively high density cloud in an inhomogeneous ISM, so that the timescale for the interaction is substantially smaller that the mass-ejection timescale.

\acknowledgments
We would like to thank an anonymous referee for detailed comments that have helped us improve the discussion in  
this paper. 
RS's contribution to the research described in this publication was
carried out at the Jet Propulsion Laboratory, California Institute of Technology, under a
contract with NASA. RS thanks NASA for financial support via a GALEX GO and ADAP award. GMC thanks
JPL for a NASA Student Independent Research Internship (SIRI).

\clearpage

\end{document}